\documentclass[reprint,aps,prd,superscriptaddress,showkeys]{revtex4-1}
\usepackage{epsfig,amsmath,natbib}

\usepackage{aas_macros}
\usepackage{amssymb}
\usepackage{amsmath}
\usepackage{dsfont}
\usepackage{hyperref}
\usepackage{color}
\hypersetup{
	colorlinks=false,
	citecolor=green
}

\def\be{\begin{equation}} 
\def\ee{\end{equation}}

\def\H2{\hbox{H$_2$}}

\def\gsim{\lower.5ex\hbox{\gtsima}} 
\def\lsim{\lower.5ex\hbox{\ltsima}} \def\gtsima{$\; \buildrel > \over 
\sim \;$} \def\ltsima{$\; \buildrel < \over \sim \;$} \def\prosima{$\; 
\buildrel \propto \over \sim \;$} \def\gsim{\lower.5ex\hbox{\gtsima}} 
\def\lsim{\lower.5ex\hbox{\ltsima}} 
\def\simgt{\lower.5ex\hbox{\gtsima}} 
\def\simlt{\lower.5ex\hbox{\ltsima}} 
\def\simpr{\lower.5ex\hbox{\prosima}} \def\la{\lsim} \def\ga{\gsim}

\def\gtsima{$\; \buildrel > \over \sim \;$} 
\def\ltsima{$\; \buildrel < \over \sim \;$} 
\def\gsim{\lower.5ex\hbox{\gtsima}} 
\def\lsim{\lower.5ex\hbox{\ltsima}} 
\def\simgt{\lower.5ex\hbox{\gtsima}} 
\def\simlt{\lower.5ex\hbox{\ltsima}} 
\def\simpr{\lower.5ex\hbox{\prosima}} 
\def\la{\lsim} 
\def\ga{\gsim}

\def\E3{{\cal E}_{\rm g}^{III}}

\begin{document}

\title{Cosmology with Minkowski functionals and moments of the weak lensing convergence field}

\author{Andrea Petri}
\email{apetri@phys.columbia.edu}
\affiliation{Department of Physics, Columbia University, New York, NY 10027, USA}
\affiliation{Physics Department, Brookhaven National Laboratory, Upton, NY 11973, USA}

\author{Zolt\'an Haiman}
\affiliation{Department of Astronomy, Columbia University, New York, NY 10027, USA}

\author{Lam Hui}
\affiliation{Department of Physics, Columbia University, New York, NY 10027, USA}
\affiliation{Department of Astronomy, Columbia University, New York, NY 10027, USA}

\author{Morgan May}
\affiliation{Physics Department, Brookhaven National Laboratory, Upton, NY 11973, USA}

\author{Jan M. Kratochvil}
\affiliation{Department of Physics, University of Miami, Coral Gables, FL 33146, USA}
\affiliation{Astrophysics and Cosmology Research Unit, University of KwaZulu-Natal, Westville, Durban, 4000, South Africa}

\date{\today}

\label{firstpage}

\begin{abstract}
  We compare the efficiency of moments and Minkowski
  functionals (MFs) in constraining the subset of cosmological parameters $(\Omega_m,w,\sigma_8)$ using
  simulated weak lensing convergence maps.  We study an analytic
  perturbative expansion of the MFs
  (\citep{Matsubara10,Munshi12}) in terms of the moments
  of the convergence field and of its spatial derivatives.  We show
  that this perturbation series breaks down on smoothing scales below
   $5^\prime$, while it shows a good degree of convergence on larger scales ($\sim 15^\prime$).
  Most of the cosmological distinguishing power is lost when the maps are smoothed on these
  larger scales. We also show that, on scales comparable to
  $1^\prime$, where the perturbation series does not converge,
  cosmological constraints obtained from the MFs are 
  approximately 1.5-2 times better
  than the ones obtained from the first few moments of the convergence
  distribution --- provided that the latter include spatial
  information, either from moments of gradients, or by combining
  multiple smoothing scales. Including either a set of these moments
  or the MFs can significantly tighten constraints on cosmological
  parameters, compared to the conventional method of using the power
  spectrum alone.
\end{abstract}


\maketitle


\section{Introduction}

Weak gravitational lensing (WL) surveys will be able to probe
cosmology with unprecedented accuracy, tightening present constraints
on some cosmological parameters
by over an order of magnitude. With recent results from the first
large WL surveys (COSMOS: \citep{Schrabback+2010,Semboloni+2011} and
CFHTLenS: \citep{Heymans+2013,Kilbinger+2013}), the question of how
to extract the maximum amount of information from weak lensing shear
and convergence maps is becoming more pressing. The power spectrum, or
equivalent two point statistics, are of unquestionable importance in
this investigation, but are inevitably incomplete if non-Gaussian
features are present. This is precisely the case when one studies weak
lensing, because gravity is non linear and generates non-Gaussian
features on small scales. The most straightforward way to characterize
non-Gaussian fields is by using higher order polyspectra, correlation
functions or moments \citep{Bernardeau1996,Jain1996,Hui99,Schneider2002,
Zaldarriaga03}.
An interesting and less explored alternative, originally proposed in the context of the 3D cosmological matter density field, is to use topological descriptors \citep{Gott86}; one kind of such descriptors are the Minkowski functionals \citep{Mecke94}. 
\citep{HikageColes} studied the effects of primordial
non-Gaussianity on the topology of large-scale structures, measuring
three-dimensional MFs on simulated density fields. \citep{Sato01} used
a next-to-leading order perturbative expansion of the MFs to derive
constraints on the matter density parameters using a limited number of
simulated weak lensing shear maps. A second early paper on MFs applied to weak lensing was written by \citep{Guimaraes02}, which used these descriptors to discern between the SCDM, OCDM and $\Lambda$CDM cosmological models. 
\citep{Munshi12} have studied a
perturbation series for the MFs in harmonic space to isolate the
contribution of different $l$ modes to each MF.  Using ray-tracing
N-body simulations, \citep{MinkJan} have recently shown that the MFs
of the convergence field can tighten constraints on dark
energy parameters by a factor
of $\approx$3 compared to using the power spectrum alone.
In this work we analyze and compare the effectiveness of the MFs and
multi-point moments of the convergence field in extracting
cosmological information. Throughout this paper, we use the term
\textit{multi-point moments} to mean moments of the convergence field
itself, as well as of its spatial derivatives. There exists a
perturbative expansion of the MFs in terms of the multi-point moments
(\citep{Matsubara10,Munshi12}). In this work, we
investigate whether this expansion converges,
both in the conventional sense of accurately reproducing the MFs, and
also in the sense of capturing the cosmological information contained
in the MFs.  The rest of this paper is organized as follows. In
\S~\ref{formalism} we give an overview of the known properties of MFs
in \S~\ref{methods} we explain the methods we used to analyze our set
of simulated maps, the results of which are summarized in
\S~\ref{results}. The results we obtained, along with a discussion,
are presented in \S~\ref{results} and \S~\ref{discussion}. Finally, in
\S~\ref{conclusion} we present our conclusions, along with a few
caveats and possible follow-ups to this project.


\section{Formalism}
\label{formalism}
\subsection{Minkowski functionals}
\label{minkowski}
\begin{figure*}
\begin{center}
\includegraphics[scale=0.6]{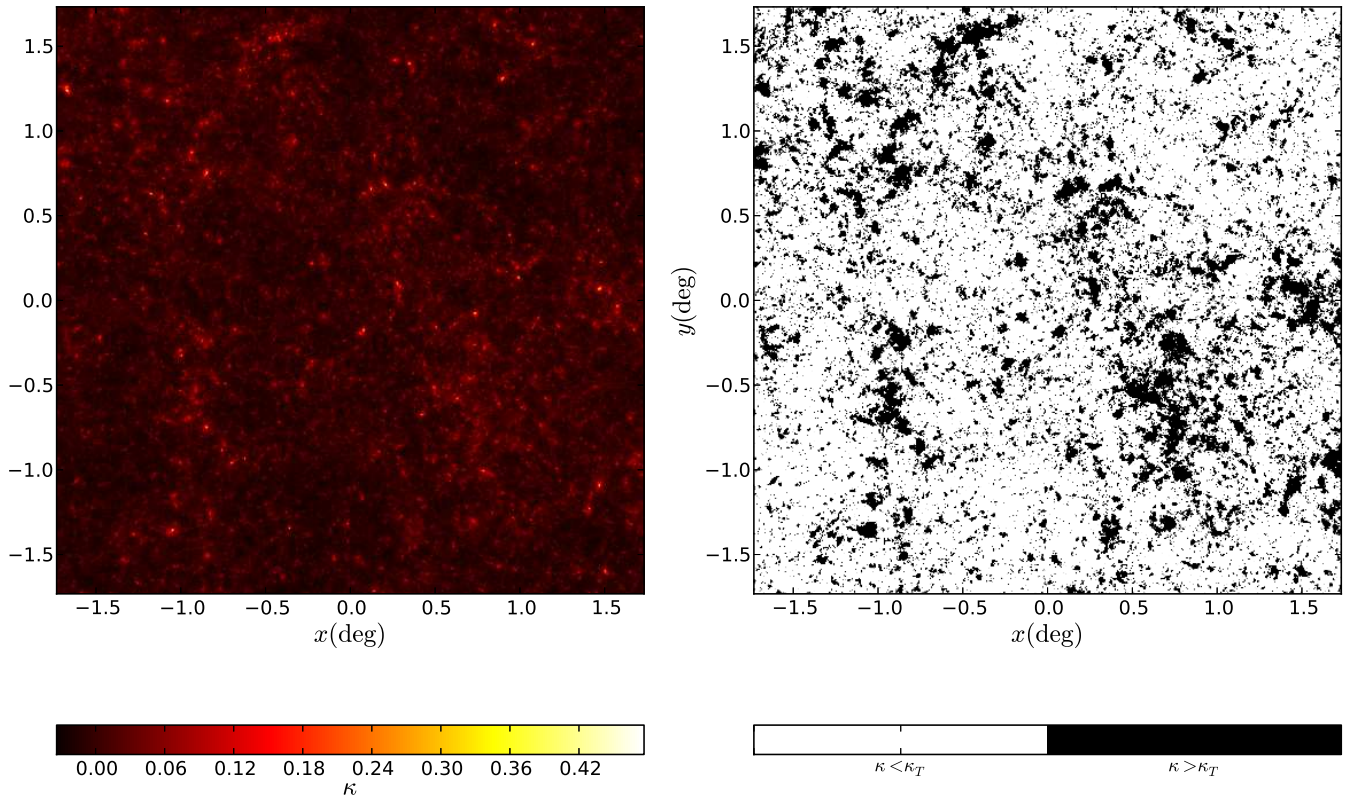}
\caption{One of our (un-smoothed) simulated convergence maps (left panel) and one of its excursion sets $\Sigma=\{\kappa>\kappa_T\}$ with $\kappa_T=0.01$ (right panel), colored in black. The MFs are measures of the area, boundary length and genus characteristic of these black regions, as a function of the threshold  $\kappa_T$.}
\label{example}
\end{center}
\end{figure*}

MFs are topological descriptors of two-dimensional random fields that have proven very useful in describing their statistical properties. Given a two-dimensional random field $\kappa(\hat{\mathbf{n}})$ (in our case the convergence field) of zero mean and variance $\langle\kappa^2\rangle=\sigma_0^2$, we can consider its excursion sets $\Sigma(\nu)=\{\kappa >\nu\sigma_0\}$ that consist of all the points at which the field exceeds a particular threshold value $\nu\sigma_0$ (see Figure \ref{example} for an illustrative example).
The three MFs $V_0(\nu), V_1(\nu)$ and $V_2(\nu)$ measure respectively the area, length of the boundary and genus characteristic of these excursion sets
\begin{equation}
\label{mf0}
V_0(\nu)=\frac{1}{A}\int_{\Sigma(\nu)}da
\end{equation}
\begin{equation}
V_1(\nu)=\frac{1}{4A}\int_{\partial\Sigma(\nu)}dl
\end{equation}
\begin{equation}
\label{mf2}
V_2(\nu)=\frac{1}{2\pi A}\int_{\partial\Sigma(\nu)}\mathcal{K}  dl
\end{equation}
where $A$ is the total area of the map, $da$ and $dl$ are the area and boundary length elements, respectively, and $\mathcal{K}$ is the curvature of the boundary. Following the notations of \citep{HikageColes} and \citep{Munshi12}, $\partial \Sigma$ in the above equations denotes the excursion set boundary. $V_0$ is a measurement of the cumulative one point PDF of the convergence field, while $V_1$ and $V_2$ contain spatial information on the excursion sets and are sensitive to $n$-point correlation properties for arbitrary high $n$. $V_2$ is related to the genus of the excursion set and is equal to the number of connected regions above the threshold (``islands''), minus the number of connected regions below the threshold (``holes'').
 For high thresholds $V_2$ is nearly equal to the number of convergence peaks \citep{Maturi+09}.
If the underlying random field is perfectly Gaussian, there is a one-to-one correspondence between the power spectrum of the field and the $V_i$ \citep{Tomita,Tomita90}.
For non-Gaussian fields, however, it is known that the MFs are sensitive to all the multi-point correlations functions, which can be used to construct a perturbative approximation \citep{Matsubara10}.

\subsection{Perturbation series}
\label{pseries}

An exhaustive analytic study of the perturbative expansion of the MFs has recently been presented by \citep{Munshi12}.  These authors have modeled in detail the harmonic-space structure of the convergence power spectrum and bispectrum, using the halo model.  In the present work, we do not consider such a fine level of detail, because we are interested only in those components of the power spectrum and the bispectrum that enter the actual perturbation expansion. In order to build a perturbative approximation to the MFs, we follow \citep{Matsubara10} and note that equations~(\ref{mf0})-(\ref{mf2}) can be conveniently expressed in terms of spatial average values of derivatives of the convergence field. Approximating the maps as flat, we introduce two coordinates $\mathbf{x}=(x,y)$ and use the notations $\alpha=\kappa/\sigma_0$, $\alpha_i=\partial_i\alpha$ and $\alpha_{ij}=\partial_i\partial_j\alpha$, with which one can write
\begin{equation}
\label{av0}
V_0(\nu)=\langle\Theta(\alpha-\nu)\rangle,
\end{equation}\begin{equation}
\label{av1}
V_1(\nu)=\frac{\pi}{8}\langle\delta(\alpha-\nu)\vert\alpha_x\vert\rangle,
\end{equation}
and
\begin{equation}
\label{av2}
V_2(\nu)=-\frac{1}{2}\langle\delta(\alpha-\nu)\delta(\alpha_x)\vert\alpha_y\vert\alpha_{xx}\rangle,
\end{equation}
where $\Theta$ is the step function and $\delta$ is the Dirac delta function. The factor $\pi/8$ in equation (\ref{av1}) comes from the fact that for isotropic fields $\langle\vert\alpha_x\vert\rangle=\frac{2}{\pi}\langle\vert\alpha\vert\rangle$, and an analogous relation can explain the factor of $-1/2$ in equation (\ref{av2}). It can be shown that these averages can be expressed in a convenient form using the variance of the convergence field, $\sigma_0$, and the variance of its gradient, $\sigma_1=\sqrt{\langle\vert\nabla\kappa\vert^2\rangle}$, as follows
\begin{equation}
\label{common}
V_k(\nu)=\frac{1}{(2\pi)^{(k+1)/2}}\frac{\omega_2}{\omega_{2-k}\omega_k}\left(\frac{\sigma_1}{\sqrt{2}\sigma_0}\right)^ke^{-\nu^2/2}v_k(\nu).
\end{equation}
Here $\omega_k$ is the measure of the solid angle in $k-1$ dimensions and hence $\omega_0=1, \omega_1=2$ and $\omega_2=\pi$.  In the Gaussian case (see \citep{Tomita,Tomita90}), the normalized MFs $v_k$ take the very simple form
\begin{equation}
\label{vgauss}
v_k^G(\nu)=H_{k-1}(\nu),
\end{equation}
where $H_k(\nu)$ is the $k$-th Hermite polynomial,
\begin{equation}
\begin{split}
& H_k(\nu)=e^{\nu^2/2}\left(-\frac{d}{d\nu}\right)^ke^{-\nu^2/2}  \\ 
& H_{-1}(\nu)=\frac{1}{2}\mathrm{erfc}\left(\frac{\nu}{\sqrt{2}}\right).
\end{split}
\end{equation}
When non--Gaussianity is present, the functions $v_k$ can be expanded in a Taylor series in powers of $\sigma_0$, whose coefficients depend on the higher-order moments of the random field. These higher moments characterize the non--Gaussianity. In particular, we can write
\begin{equation}
\label{expansion}
v_k(\nu)=v_k^G(\nu)+\sum_{m=1}^{\infty}v_k^{(m)}\sigma_0^m.
\end{equation}
The series coefficients $v_k^{(m)}$ will, in general, contain the information on the higher--order moments of the distribution, such as the skewness and the kurtosis.  These coefficients take the general form
\begin{equation}
v_k^{(m)}(\nu)=\sum_{n=0}^{3m+1} c^{(m)}_{kn} H_n(\nu).
\end{equation}
As an example, the first term ($m=1$) in the expansion (eq.~\ref{expansion}) is given by \citep{Matsubara10} 
\begin{equation}
c^{(1)}_{k,k+2}=\frac{S_0}{6} ,\,\,\,\,c^{(1)}_{k,k}=-\frac{kS_1}{4} ,\,\,\,\, c^{(1)}_{k,k-2}=-\frac{k(k-1)S_2}{4},
\end{equation}
where
\begin{equation}
\label{skew}
S_0=\frac{\langle\kappa^3\rangle}{\sigma_0^4}, \,\,\,\, S_1=\frac{\langle\kappa^2\nabla^2\kappa\rangle}{\sigma_0^2\sigma_1^2}, \,\,\,\, S_2=\frac{2\langle\vert\nabla\kappa\vert^2\nabla^2\kappa\rangle}{\sigma_1^4}.
\end{equation}
Note that this term is determined entirely by the three skewness parameters (\ref{skew}). The next to leading order corrections ($m=2$) are similarly given by a weighted sum of the four connected kurtosis parameters. 
\begin{equation}
\label{kurt}
\begin{split}
& K_0=\frac{\langle\kappa^4\rangle_c}{\sigma_0^6}, \,\,\,\, K_1=\frac{\langle\kappa^3\nabla^2\kappa\rangle_c}{\sigma_0^4\sigma_1^2}, \\
& K_2=\frac{\langle\kappa\vert\nabla\kappa\vert^2\nabla^2\kappa\rangle_c}{\sigma_0^2\sigma_1^4} , \,\,\,\, K_3=\frac{\langle\vert\nabla\kappa\vert^4\rangle_c}{\sigma_0^2\sigma_1^4}.
\end{split}
\end{equation}
In this work by ''connected'' we mean the non-Gaussian components of the moments: for Gaussian fields we have that $\langle\kappa^4\rangle_G=3\langle\kappa^2\rangle^2_G$, so that the connected part $\langle\kappa^4\rangle_c=\langle\kappa^4\rangle-3\langle\kappa^2\rangle^2$ is a measure of the second order non-Gaussianity. In the Gaussian case the skewness parameters are zero, and hence they coincide with their connected part.  
The expressions for these next-to-leading-order corrections are
lengthy and do not add anything to the present discussion; we refer
the interested reader to \citep{Matsubara10} for the complete set of
equations.   However, it is worth pointing out a relevant property of
the expansion series. Since $V_0$ does not contain any spatial or
morphological information about the field, it makes sense that its
perturbation series, at all orders, will be entirely determined in
terms of the one point cumulants $\langle\kappa^n\rangle_c$.   In
contrast, the moments with derivatives, which contain spatial
information, will appear in the expansion of both of the other two
MFs. In order to have any hope that the series
convergences, it must be that the higher order cumulants become
smaller and smaller as $n$ grows, i.e. it is necessary that
$\langle\kappa^n\rangle_c/\sigma_0^n\rightarrow 0$ as $n$ grows.  When
the departure from a Gaussian field is significant, this is generally not the case --- the series then does not converge, and the MFs do not admit a perturbative expansion. In the case of temperature maps of the cosmic microwave background (CMB), which are almost Gaussian, it has been shown (e.g. \citep{Matsubara10,Lim12}) 
that the series converges accurately, also because the simple Gaussian approximation is good enough. Given the much larger non-Gaussianities in the WL convergence field, it is not guaranteed that the series will converge when applied to WL.  A goal of the present study is to quantify how well (or not) the WL series converges.


\section{Methods}
\label{methods}

\subsection{N-body simulations and ray-tracing}
\label{simulations}
In order to understand if the weak lensing MFs admit a perturbative expansion in terms of multi-point moments, we must evaluate the series numerically and compare it to the actual MFs measured from simulated maps. For this purpose, we use a large suite of simulated convergence maps (see \citep{MinkJan}), generated with a two-dimensional ray-tracing algorithm (see \citep{PeaksJan}), for a number of different combinations of the three cosmological parameters $(\Omega_m,w,\sigma_8)$, referring respectively to the matter density, the equation of state of dark energy and the normalized amplitude of the initial density fluctuations, as summarized in Table \ref{simsuite}.
\begin{table}
\begin{center}
\begin{tabular}{|c|c|c|c|c|} \hline
Description & $\Omega_m$ & $w$ & $\sigma_8$ & Number of simulations \\ \hline
Fiducial & 0.26 &-1.0 &0.798 & 45 \\
Auxiliary & 0.26 &-1.0 &0.798 & 5 \\
Low $\Omega_m$ & 0.23 &-1.0 &0.798 & 5 \\
High $\Omega_m$ & 0.29 &-1.0 &0.798 & 5 \\
Low $w$ & 0.26 &-1.2 &0.798 & 5 \\
High $w$ & 0.26 &-0.8 &0.798 & 5 \\
Low $\sigma_8$ & 0.26 &-1.0 &0.750 & 5 \\
High $\sigma_8$ & 0.26 &-1.0 &0.850 & 5 \\ \hline
\end{tabular}
\end{center}
\caption{Available cosmologies for ray-tracing simulations; all models describe a spatially flat universe with $\Omega_\Lambda+\Omega_m=1$.}
\label{simsuite}
\end{table}
For each choice of cosmological parameters we have $R=1000$ (pseudo-)independent realizations of the convergence field for each of three fixed source galaxy redshifts $z_s=1,1.5$ and 2. Each realization is a flat 12\,deg$^2$ map with a pixel resolution of $A_p=(0.1\,\mathrm{arcmin})^2$ and $N_p=2,048$ pixels per side. We call these realizations (pseudo-)independent because they are drawn from the same set of simulations, rotating the lens planes from which the ray tracing is performed. We add galaxy shape noise, $\kappa_{noise}$, to each of the realizations, following a conventional approach (see for example \citep{MinkJan}) to model this noise as a white Gaussian noise with real space correlation function
\begin{equation}
\langle\kappa_{noise}(\mathbf{x}_1)\kappa_{noise}(\mathbf{x}_2)\rangle=\frac{\sigma^2_\gamma(z_s)}{n_{gal}A_p}\delta^K_{\mathbf{x}_1,\mathbf{x}_2}.
\end{equation}
Here $n_{gal}$ is the number density of source galaxies, which we assume to be $n_{gal}=15\mathrm{arcmin}^{-2}$, $\delta^K$ is the Kronecker delta tensor and each pixel $\mathbf{x}$ is specified by its two coordinates $(x,y)$. We assume that this noise has an amplitude equal to that of one component of the shear, and so has an r.m.s. value (see \citep{SongKnox}) 
\begin{equation}
\sigma_\gamma(z_s)=0.15 + 0.035z_s.
\end{equation}
After adding the noise, we smooth each map with a Gaussian window function of angular size $\theta_G$ and kernel
\begin{equation}
W_{\theta_G}(x,y)=\frac{1}{\pi\theta_G^2}\exp\left[{-\left(\frac{x^2+y^2}{\theta_G^2}\right)}\right]
\end{equation}
Throughout this paper, we refer to the parameter $\theta_G$ as the smoothing scale of our maps; for reference, the maps are stored in FITS format and the input is processed with the CFITSIO library \citep{FITS}; the smoothing is performed in Fourier space for computational time speedup using the FFTW3 C library \citep{FFTW05}.

\subsection{Observables from simulated lensing maps}
\label{obsmeasurements}
For each map, we measure a set of multi--point moments and MFs with linear size threshold bins.
We limit our perturbation series expansion to order $\sigma_0^2$ and, motivated by equations (\ref{skew}), (\ref{kurt}), we measure the nine moments $O^{M,r}_i=(\sigma_0^2,\sigma_1^2,S_0,S_1,S_2,K_0,K_1,K_2,K_3)_r$ for each realization $r$; gradients are calculated with first order finite difference looking at the values of the nearest neighboring pixels, assuming periodic boundary conditions.  
For the direct measurement of the MFs we again follow \citep{MinkJan} and express the spatial averages (\ref{av0}),(\ref{av1}) and (\ref{av2}) as integrals over the map planes 
\begin{equation}
\label{v0meas}
V_0(\nu)=\frac{1}{A}\int_A\Theta(\kappa(\mathbf{x})-\nu)dxdy,
\end{equation}
\begin{equation}
\label{v1meas}
V_1(\nu)=\frac{1}{4A}\int_A\delta(\kappa(\mathbf{x})-\nu)\sqrt{\kappa_x^2+\kappa_y^2}dxdy,
\end{equation}
\begin{equation}
\label{v2meas}
V_2(\nu)=\frac{1}{2\pi A}\int_A\delta(\kappa(\mathbf{x})-\nu)\frac{2\kappa_x\kappa_y\kappa_{xy}-\kappa_x^2\kappa_{yy}-\kappa_y^2\kappa_{xx}}{\kappa_x^2+\kappa_y^2}dxdy.
\end{equation}
In equations (\ref{v0meas})-(\ref{v2meas}) the integrals are calculated as discrete sums over pixels and the derivatives are evaluated via finite difference with periodic boundary conditions. For large smoothing scales $\theta_G$, the denominator in the integrand of expression (\ref{v2meas}) can vanish on some of the pixels; this is an issue especially for large smoothing scales, when neighboring pixels can acquire similar values, making gradients vanish. 
 To deal with this, when this happens, we take the explicit limit $\vert\nabla\kappa\vert\rightarrow 0$ with $\kappa_x=\kappa_y$ and replace the integrand with the expression $\kappa_{xy}-(\kappa_{xx}+\kappa_{yy})/2$. The direction in which to take this limit is arbitrary, but the difference between one choice and the other is of relative order $1/N_p^2$ and hence can be safely neglected. To approximate the $\delta$ functions we use discrete binning for the thresholds $\nu_j$ and we divide the interval $\kappa=\nu\sigma_0 \in [-\vert\kappa_{min}\vert,\vert\kappa_{min}\vert]$ in $N_{bins}$ parts of equal size. This discretization choice leads to rounding errors for the MFs, which have been investigated by \citep{MinkJan}. We then save these binned MFs in a vector of size $3N_{bins}$, $O^{MF,r}_i=[V_0(\nu_j),V_1(\nu_j),V_2(\nu_j)]^r$ for each realization $r$. Our complete set of descriptors is then a $9+3N_{bins}$ dimensional vector $O_i^r=(O^{M,r},O^{MF,r})$. We measure this entire set of descriptors for each of the 1000 realizations for each cosmological model, using 100 cores at the BNL Astro Cluster. Parallelization is implemented with the MPI library \citep{mpi}. 

\subsection{Forecasting cosmology constraints}
\label{stats}
To measure the information content (i.e. the constraining power) that each of our descriptors $O_i$
carries, we want to calculate the error on each of the cosmological parameters $p_\alpha=(\Omega_m,w,\sigma_8)$ that results from using a particular set of descriptors. To begin, we measure the averages $\langle{O}_i\rangle$ and the covariance matrix $C_{ij}$ of our descriptors using the 1000 realizations available,
\begin{equation}
\langle O_i\rangle=\frac{1}{R}\sum_{r=1}^RO_i^r,
\end{equation}
and
\begin{equation}
\label{covest}
C_{ij}=\frac{1}{R-1}\sum_{r=1}^R(O_i^r-\langle O_i\rangle)(O_j^r-\langle O_j\rangle).
\end{equation}
Before using this information to fit for the cosmological parameters, we can ask ourselves how well our set of descriptors $O_i$ is able to discriminate between two different cosmological models.  We have quantified this difference using a $\chi^2$ analysis, and we computed the quantity
\begin{equation}
\label{dchi2}
\Delta \chi^2_{ff^\prime}= (\langle O_i\rangle^f-\langle O_i\rangle)^{f^\prime}(C^f)^{-1}_{ij}(\langle O_j\rangle^f-\langle O_j\rangle^{f^\prime}).
\end{equation}
Note that repeated indices $i$ and $j$ are summed over, and that the superscripts $r$ and $f$ in equations (\ref{covest}) and (\ref{dchi2}) have different meanings: in the former case we refer to the superscript as the particular realization we are analyzing, in the latter as the particular cosmological model considered. The quantity $\Delta \chi^2_{ff^\prime}$ measures how well a set of descriptors $O_i$ is able to discriminate between two different cosmological models, typically the fiducial {\em v.s.} one of the alternative models in Table~\ref{simsuite}. We use the same set of descriptors to find the best-fit cosmological parameters $p_\alpha^r=p^0_\alpha+\delta p_\alpha^r$ [with $p^0_\alpha=(0.26,-1.0,0.798)$] for each realization; this allows us to compute the parameter covariance matrix $P_{\alpha\beta}=\langle\delta p_\alpha\delta p_\beta\rangle$. 
We use a linear interpolation between the fiducial and the alternative cosmologies to model 
the dependence of each descriptor on the cosmological parameters, 
\begin{equation}
\label{interp}
O_i(p)=\langle O_i(p^0)\rangle+X_{i\alpha}\delta p_\alpha
\end{equation}
with $X_{i\alpha}=\partial \langle O_i\rangle/\partial p_\alpha$ is evaluated as a forward finite difference derivative.
In each realization (of the fiducial cosmology), we find the best-fit cosmological parameters by minimizing the $\chi^2$,
\begin{equation}
\label{dchi2fit}
\chi^2_r(p)=[O_i^r-O_i(p)]C^{-1}_{ij}(p^0)[O_j^r- O_j(p)]
\end{equation}  
with respect to $p$ (note that $p$ is a vector of the three cosmological parameters). We want to stress the difference in meaning by the two $\chi^2$ statistics (\ref{dchi2}),(\ref{dchi2fit}): the former is a measure of the cosmological distinguishing power of a particular set of descriptors, the latter is a function of $p$, which minimum represents the best fit parameter values for an individual realization. With the linear interpolation (\ref{interp}), this minimization can be done by solving
$\partial \chi^2_r(p)/\partial p = 0$ analytically. This minimization condition translates to
\begin{equation}
\label{tosolve}
X_{i\alpha}C^{-1}_{ij}(p^0)[O_j^r-\langle O_j(p^0)\rangle-X_{j\beta}\delta p_\beta^r]=0
\end{equation}  
The solution $\delta p_\alpha$ to this linear system is given in terms of the differences $\delta O_i^r=O_i^r-\langle O_i(p^0)\rangle$ and inverse matrices,
\begin{equation}
\label{fitted}
\delta p^r_\alpha=(X_i C^{-1}_{ij}X_j)^{-1}_{\alpha\beta}(X_{k\beta}C^{-1}_{kl}\delta O_l^r)
\end{equation}
We can also calculate the parameter covariance matrix $P$ analytically,
in terms of the covariance matrix of the descriptors, using equation (\ref{fitted}) and the fact that $\langle\delta O_i \delta O_j\rangle=C_{ij}$; matrix manipulations give us
\begin{equation}
\label{covanalytical}
P_{\alpha\beta}= (X_i C^{-1}_{ij}X_j)^{-1}_{\alpha\beta}.
\end{equation}
The final result of this calculation is the inverse of the familiar
Fisher matrix $F_{\alpha\beta}=(X_i C^{-1}_{ij}X_j)_{\alpha\beta}$,
except neglecting the dependence of the (co)variances on cosmology
(\citep{TegmarkFisher,PeaksXiuyuan}); we should stress that this statement holds only because we chose to perform the fitting of the cosmological parameters linearly interpolating our observables set between different cosmological models.
We should also note that the marginalized errors calculated from this
analytical covariance matrix $\Delta p_\alpha=\sqrt{P_{\alpha\alpha}}$
correspond to ``one sigma'', or $68.4\%$ confidence level only if the
fluctuations of the observables between realizations are Gaussian. We
have no reasons apriori that this would be the case for the weak
lensing MFs and moments.  On the other hand, the above approach
directly yields the full likelihood analytically (since we have the
best-fit parameters for each realization), and allows us to go beyond
the Fisher matrix and relax this Gaussian assumption.
Further details and discussion of this point will be included in
\S~\ref{infcontent} below. 
There is an additional complication: as we will clarify in \S~\ref{robustness}, we have reasons to think that this picture is oversimplified and that the parameter error bars that one gets from equation (\ref{covanalytical}) are underestimated when the descriptor set (and in particular the number of thresholds for the MF, $N_{bins}$) is too large. To correct this underestimation, we take advantage of the fact that we have at our disposal two different set of maps for the fiducial cosmology, generated respectively from 5 and 45 simulations. We use one set of maps, $A$, to measure the covariance matrix which we use to construct the $\chi^2$ in equation (\ref{dchi2fit}); we call this covariance matrix $C^A$. We use the other set of maps, $B$, to fit the cosmological parameters; that is to say in equation (\ref{dchi2fit}) the observables $O_i^r$ are measured from the set of maps $B$ so that $\langle\delta O_i\delta O_j\rangle=C^B_{ij}$ is the covariance matrix measured from set $B$. This leads to a more complicated expression for the parameter covariance, which can be generalized as 
\begin{equation}
\label{covanalyticalmod}
P^{AB}_{\alpha\beta}=(X\cdot Y_{C^A})^{-1}_{\alpha\gamma}(X\cdot Y_{C^A})_{\beta\delta}^{-1}(Y_{C^A})_{i\gamma}(Y_{C^A})_{j\delta}C^B_{ij},
\end{equation}
with the convenient notation $Y_{C}=C^{-1}X$. As pointed out by \citep{MinkJan}, the covariance matrices measured from different set of maps (obtained with 5 and 45 simulations) are not the same and equation (\ref{covanalyticalmod}) can lead to quite different (but more accurate) results than the simpler equation (\ref{covanalytical}). In the present discussion, we didn't mention the fact that equations (\ref{dchi2}) and (\ref{covanalytical}) need to be corrected by a multiplying factor of $(R-3N_{bins}-2)/(R-1)$ due to the fact that the estimator that we use for the inverse of the covariance matrix, $C^{-1}_{ij}$, is biased (see \citep{Hartlap07,Anderson03}); for our set of MFs binned with $N_{bins}=100$ this accounts for a 10\% correction, which is small enough not to affect any of our conclusions below. The finding that equation (\ref{covanalytical}) underestimates error bars is not due to the fact that it misses the correction factor, but to the fact that the correlation between the observables vector and the covariance matrix forces us to use different set of maps, and in particular equation (\ref{covanalyticalmod}) to obtain the correct sized error bars.
Throughout the discussions below, the errors are calculated as $\Delta p_\alpha=\sqrt{P^{AB}_{\alpha\alpha}}$ where $A$ and $B$ refer to the sets of fiducial maps obtained respectively from 5 and 45 simulations; the errors are scaled by the constant factor $1/\sqrt{N_{LSST}}$, where $N_{LSST}=1600$ is the approximate ratio of the solid angle covered by the LSST survey
 ($\approx$20,000 deg$^2$; \citep{LSST2.0}) to the area of one of our simulated maps
($\approx$12 deg$^2$).


\section{Results}
\label{results}

\subsection{Numerical MF measurement accuracy}
\label{accuracy}
To check the accuracy of our code, we have created 1,000 realizations of a Gaussian random field, with the same size as our
weak lensing maps (see \citep{MinkJan})
and measured the three MFs with the procedure in \S~\ref{obsmeasurements}.  We have used $N_{bins}=100$, choice that will be justified in \S~\ref{robustness}

\begin{figure*}
\begin{center}
\includegraphics[scale=0.3]{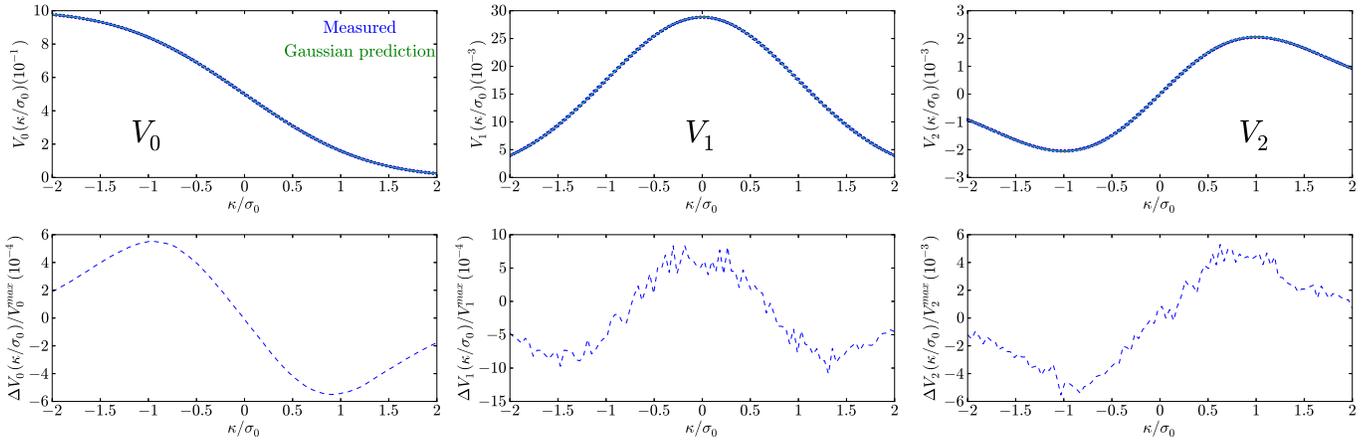}
\caption{Comparison between the  Minkowski functionals measured numerically on the simulated maps, and the analytical predictions from equation (\ref{vgauss}), for 1000 realizations of a Gaussian random field (with the same size and pixel resolution as our convergence maps). The figure shows a direct comparison (top panel) and the fractional errors  $\delta V/V_{max}$ (bottom panel) for the three MFs ($V_0, V_1, V_2$ from left to right).}
\label{gausscomp}
\end{center}
\end{figure*}

The results are shown, together with the analytic expectations from equations (\ref{common}) and (\ref{vgauss}), in Figure \ref{gausscomp}.
As this figure shows, our code measures the MFs for a Gaussian random field highly accurately, with relative residuals smaller than one part in $10^3$.   Although \citep{MinkJan} achieved somewhat better accuracy, the accuracies in Figure \ref{gausscomp} are sufficient for our purposes, and will not influence our conclusions on the convergence of the perturbation series. We will discuss this point further in \S~\ref{discussion} below.

\subsection{Perturbation series convergence}
\label{seriesconv}
\begin{figure*}
\begin{center}
\includegraphics[scale=0.3]{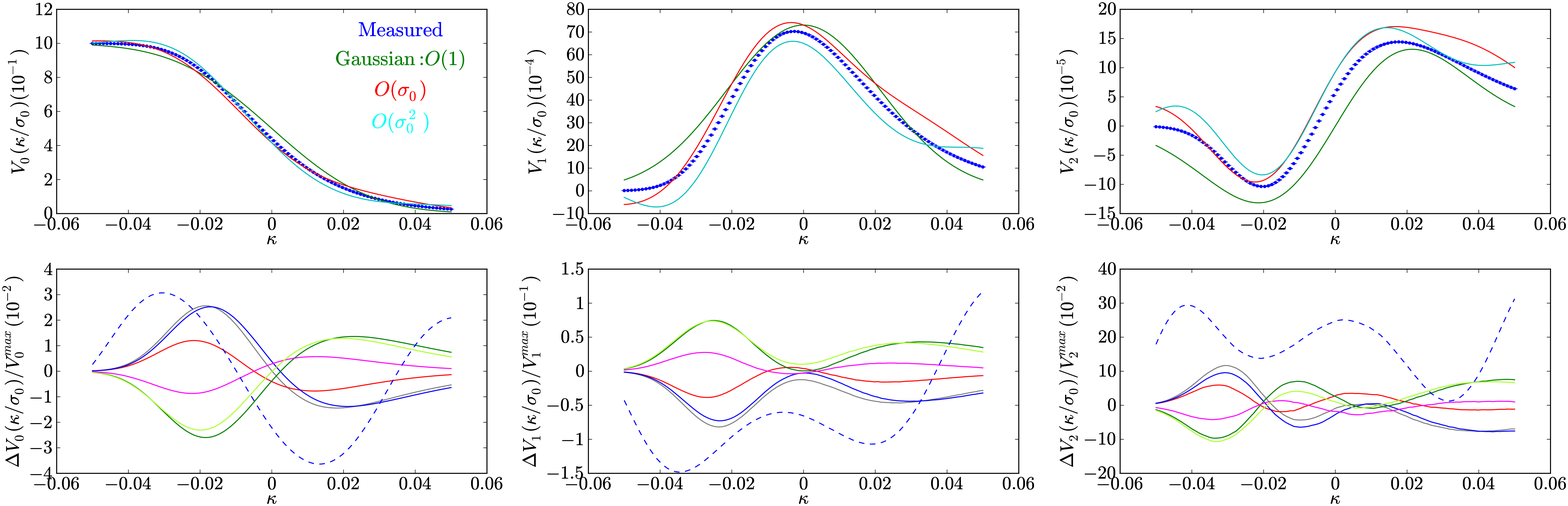}
\includegraphics[scale=0.3]{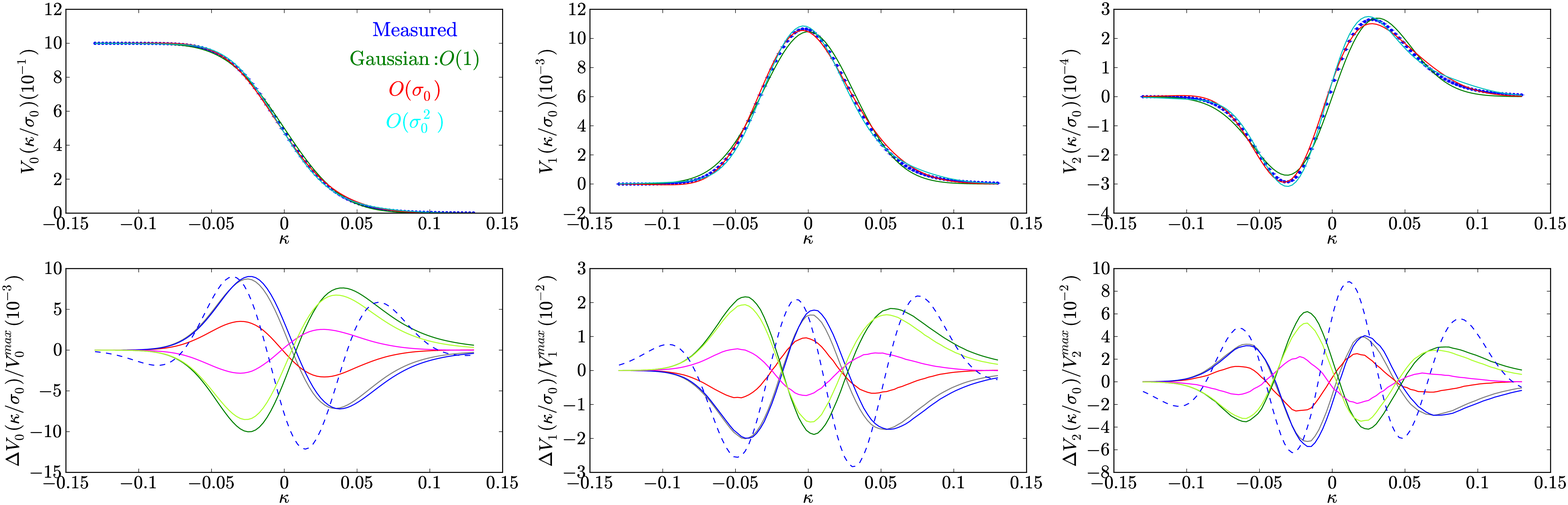}
\caption{Tests of the accuracy of the perturbation series expansion
  for the MFs.  In each panel, we show the MFs measured numerically
  from the simulated WL maps (blue points) compared with the
  prediction from the perturbation series at Gaussian (green),
  $\sigma_0$ (red) and $\sigma_0^2$ (cyan) order.  In each case, the
  observables were averaged over 1000 realizations of a 12 deg$^2$ WL
  map, with source redshift $z_s=2$, smoothed on a scale
  $\theta_G=1^\prime$.  The top pair of rows use the noiseless maps,
  while in the bottom pair of rows shape noise was added.  In each
  pair, the upper panel compares the MFs directly, and the lower
  panels shows the residuals $\delta V_k$ (dashed blue line) between the measured MFs and the order $\sigma_0^2$ perturbation series, together with the
  difference in MFs between the fiducial cosmology and cosmological models with different
  $(\Omega_m,w,\sigma_8)$ (solid lines). The colors refer to $\Omega_m=0.23$ (grey), $w=-0.8$ (red), $\sigma_8=0.75$ (blue), $\Omega_m=0.29$ (yellow),  
$w=-1.2$ (magenta) and $\sigma_8=0.85$ (green).
  Note that the scale on the y axis is different in each row, and the
  noisy maps have smaller residuals, as expected from a field that is
  closer to Gaussian. }
\label{comparepert}
\end{center}
\end{figure*}

\begin{figure*}
\begin{center}
\includegraphics[scale=0.3]{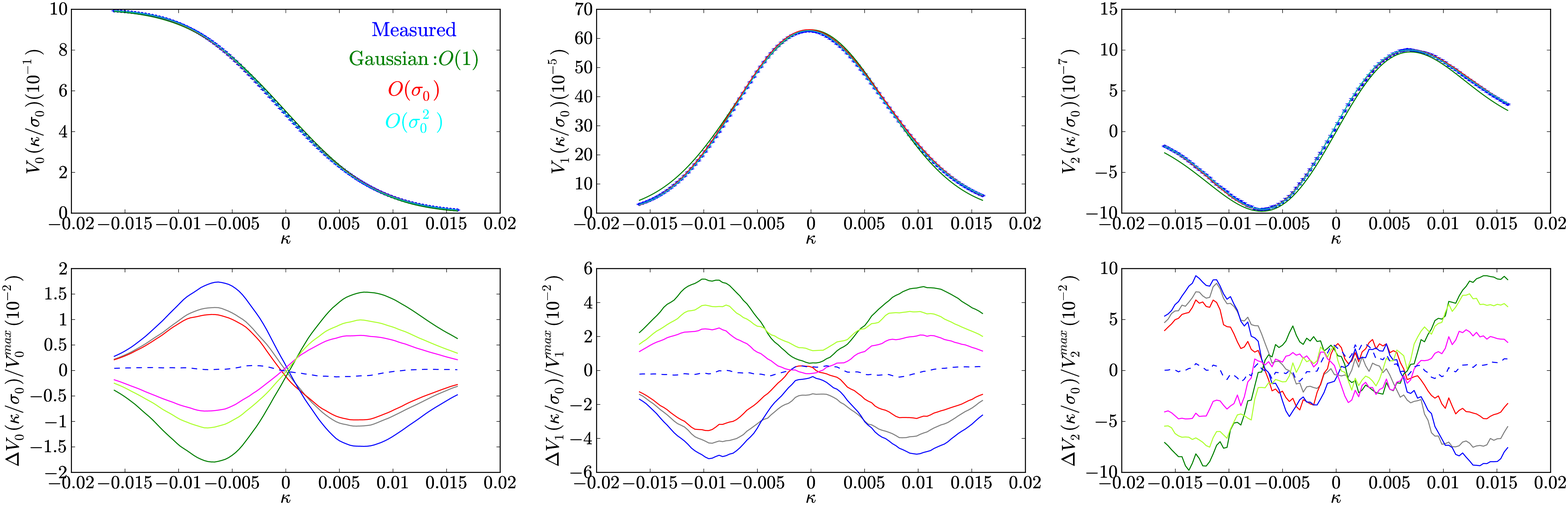}
\includegraphics[scale=0.3]{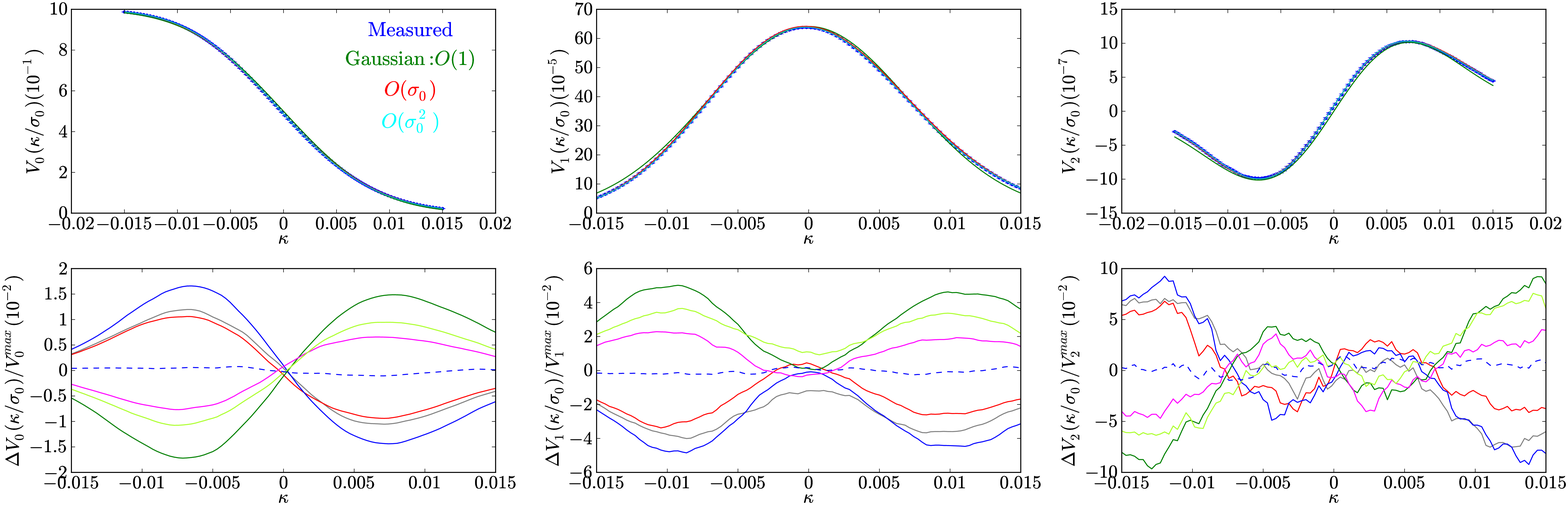}
\caption{Tests of the accuracy of the perturbation series expansion
  for the MFs, as in Figure~\ref{comparepert}, except for larger
  smoothing scales.  The top pair of panels show results from
  noiseless maps, with the smoothing scale increased to
  $\theta_G=15^\prime$; the bottom panel shows results from maps with
  shape noise added, smoothed on a scale $\theta_G=15^\prime$. The color code is exactly the same as Figure \ref{comparepert}.}
\label{comparepertbetter}
\end{center}
\end{figure*}

\begin{figure*}
\begin{center}
\includegraphics[scale=0.5]{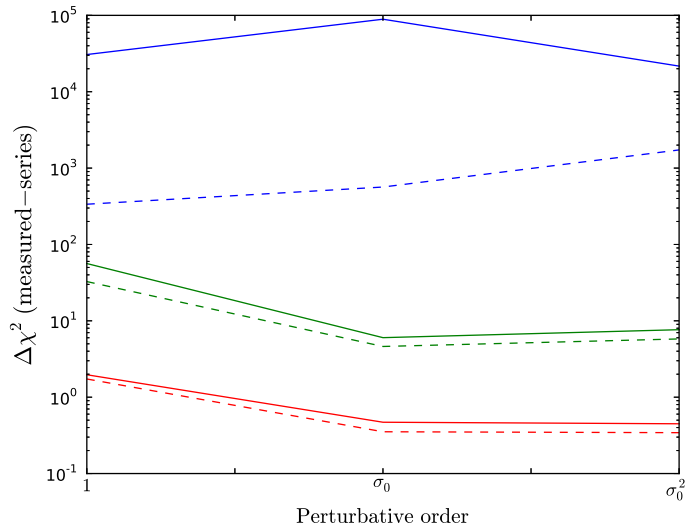}
\caption{$\Delta \chi^2$ difference between the measured Minkowski functionals and their perturbative approximation, as a function of the perturbative order considered in powers of $\sigma_0$; we consider the noiseless (solid) and noisy (dashed) cases with $n_{gal}=15\mathrm{arcmin}^{-2}$ and $z_s=2$ for smoothing scales $\theta_G=1^\prime$(blue),$5^\prime$(green) and $15^\prime$(red). The calculations are performed with $N_{bins}=100$.}
\label{chi2convergence}
\end{center}
\end{figure*}

We next test the  convergence of the perturbation series in equation~(\ref{expansion}) up to order $\sigma_0^2$, i.e. up to the contributions containing the fourth order cumulants. We compared the residuals $\delta V_k=V_k^{meas}-V_k^{(2)}$ with the differences in $V_k$ that we obtain when varying the cosmological parameters $(\Omega_m,w,\sigma_8)$; we again used $N_{bins}=100$. 
The MFs from the analytical series expansion, $V_k^{(2)}$, are
calculated by measuring the moments for each of the 1000 realizations,
taking the average, and then using equations (\ref{common}) to
(\ref{kurt}). The results are displayed in Figures \ref{comparepert}
and \ref{comparepertbetter}, for both the noisy and the noiseless
maps, and for different smoothing scales.  One expects that for a
sufficiently large smoothing scale, the field becomes close to
Gaussian, and the perturbation series is convergent. We investigated this issue in Figure \ref{chi2convergence}, where we show the $\Delta \chi^2$ difference (calculated with equation (\ref{dchi2})) between the measured and perturbative expanded MFs as a function of the perturbative order and smoothing scale used. If the series is convergent, an
important question to ask is if the MFs are still
effective in distinguishing different cosmological models.  To answer
this question, we computed the $\Delta \chi^2$ values as in equation
(\ref{dchi2}) for each of the smoothing scales we tested, and for
three deviations from the fiducial cosmology, summarizing the results
in Table \ref{chi2}. We used the same $\Delta \chi^2$ statistic to quantify the convergence of the perturbation series, and see how this compares to the difference in cosmological models. 

\begin{table*}
\begin{center}
\begin{tabular}{|c|c|c|}\hline
$\theta_G$(arcmin) & \multicolumn{2}{|c|}{$\Delta \chi^2$} \\ \hline
&  \textbf{Noiseless} & \textbf{Noisy} \\ \hline
&\multicolumn{2}{|c|}{\textbf{Analytical - measurement comparison ($O(\sigma_0^2)$)}} \\ \hline
1 & $2.16\cdot 10^4$ & $1.73\cdot 10^3$  \\
5 &  7.65 & 5.79  \\
15 &  0.45 &  0.34\\ \hline
&\multicolumn{2}{|c|}{\textbf{High $\Omega_m$}} \\ \hline
1 & 32.45 & 9.32 \\
5 & 5.48 & 3.86 \\
15 & 1.47 & 1.18 \\ \hline
&\multicolumn{2}{|c|}{\textbf{High $w$}} \\ \hline
1 &  20.60 &  2.67 \\
5  & 2.37 & 1.59 \\
15 & 1.47 & 1.13 \\ \hline
&\multicolumn{2}{|c|}{\textbf{High $\sigma_8$}} \\ \hline
1 &  21.06 & 12.84 \\
5  & 6.42  & 5.14 \\
15 & 1.47 & 1.44 \\ \hline
\end{tabular}
\caption{Distinguishing power between the fiducial model and three of its variants using the Minkowski functionals, measured by the $\Delta \chi^2$ defined in equation (\ref{dchi2}); the calculations are performed with $N_{bins}=100$. In the first row we show the $\Delta \chi^2$ difference between the second order perturbation series and the actually measured MFs in the fiducial cosmology.}
\label{chi2}
\end{center}
\end{table*}

\subsection{Information content and constraining power}
\label{infcontent}
We calculated the marginalized errors in the cosmological parameters analytically using equation (\ref{covanalyticalmod}) for various sets of descriptors, in order to distinguish the constraining power of the MFs and the moments. We chose again $N_{bins}=100$ to obtain realistic constraints: as already stressed this choice will be justified later in \S~\ref{robustness}.
The results are summarized in Table \ref{momcontributions} and will be discussed in \S~\ref{discussion} below
\begin{table*}
\begin{center}
\begin{tabular}{|c|c|c|c|} \hline
Descriptors &$\Delta \Omega_m$ & $\Delta w$ & $\Delta \sigma_8$ \\ \hline
\multicolumn{4}{|c|}{\textbf{One point moments}} \\ \hline
Unconnected ($\langle\kappa^2\rangle,\langle\kappa^3\rangle,\langle\kappa^4\rangle$) & 0.0066  &0.035 &0.0069 \\
Connected ($(\sigma_0^2,S_0,K_0)=\langle\kappa^2\rangle,\langle\kappa^3\rangle,\langle\kappa^4\rangle_c$) & 0.0051 & 0.025 & 0.0053 \\ \hline
\multicolumn{4}{|c|}{\textbf{All moments (connected)}} \\ \hline
Add $\sigma_1^2=\langle\vert\nabla\kappa\vert^2\rangle$ & 0.0017 & 0.0089 & 0.0023 \\ \hline
Add $S_1=\langle\kappa^2\nabla^2\kappa\rangle$ & 0.0016 & 0.0088 & 0.0022 \\
Add $S_2=\langle\vert\nabla\kappa\vert^2\nabla^2\kappa\rangle$& 0.0016 & 0.0082 & 0.0021 \\ \hline
Add $K_1=\langle\kappa^3\nabla^2\kappa\rangle_c$ & 0.0016& 0.0082 & 0.0021 \\
Add $K_2=\langle\kappa\vert\nabla\kappa\vert^2\nabla^2\kappa\rangle_c$ & 0.0016 & 0.0082 & 0.0021 \\
Add $K_3=\langle\vert\nabla\kappa\vert^4\rangle_c$ & 0.0015 & 0.0081 & 0.0020 \\ \hline
\multicolumn{4}{|c|}{\textbf{Smoothing scale combination}} \\ \hline
$(\sigma_0^2,S_0)\times(1^\prime,3^\prime)$ & 0.0020 & 0.012 & 0.0025 \\ 
$(\sigma_0^2,\sigma_1^2,S_0)\times(1^\prime,3^\prime)$ & 0.0016 & 0.0092 & 0.0021 \\ 
$(\sigma_0^2,S_0,K_0)\times(1^\prime,3^\prime)$ & 0.0019 & 0.010 & 0.0024 \\ 
$(\sigma_0^2,S_0,K_0)\times(1^\prime,3^\prime,5^\prime)$ &0.0017 &0.0090 &0.0022 \\ \hline
\multicolumn{4}{|c|}{\textbf{Minkowski functionals}} \\ \hline
$V_0$  & 0.0015 &0.0073 &0.0019 \\ 
$V_1$  & 0.0013 &0.0065 &0.0018 \\ 
$V_2$  & 0.0014 &0.0067&0.0020\\ 
$V_0+V_1+V_2$  & 0.00096 &0.0052&0.0014 \\ \hline
\multicolumn{4}{|c|}{\textbf{Minkowski functionals + All moments}} \\ \hline
& 0.00096&0.0051&0.0014 \\ \hline
\end{tabular}
\end{center}
\caption{The marginalized errors on the cosmological parameters, calculated 
  from $P^{AB}_{\alpha\beta}$ (equation (\ref{covanalyticalmod})), and using
  various sets of observables.  In the top section of the table, we 
  start by considering only the three one--point moments (unconnected and connected), 
  and then we add the connected derivative moments one by one to the ensemble. 
  In the middle section, we consider the effect of combining smoothing scales (this is only done is this section; all the other sections in this table refer to a single smoothing scale $\theta_G=1^\prime$).
  The bottom section shows the constraints from the MFs, either alone, or combined 
  with the moments.  In each case, the source redshift is fixed at $z_s=2$ and
  the maps are smoothed on an angular scale $\theta_G=1^\prime$. 
  Galaxy shape noise with $n_{gal}=15\,\mathrm{arcmin}^{-2}$ has been added.}
\label{momcontributions}
\end{table*}

\begin{figure*}
\begin{center}
\includegraphics[scale=0.3]{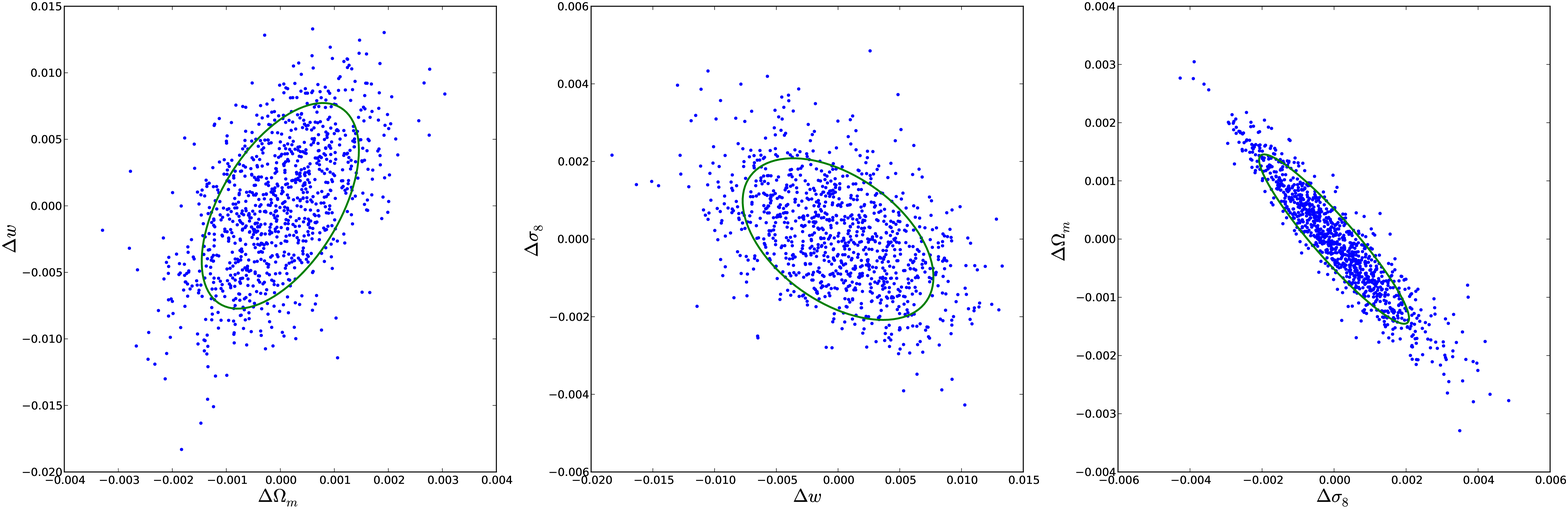}
\includegraphics[scale=0.3]{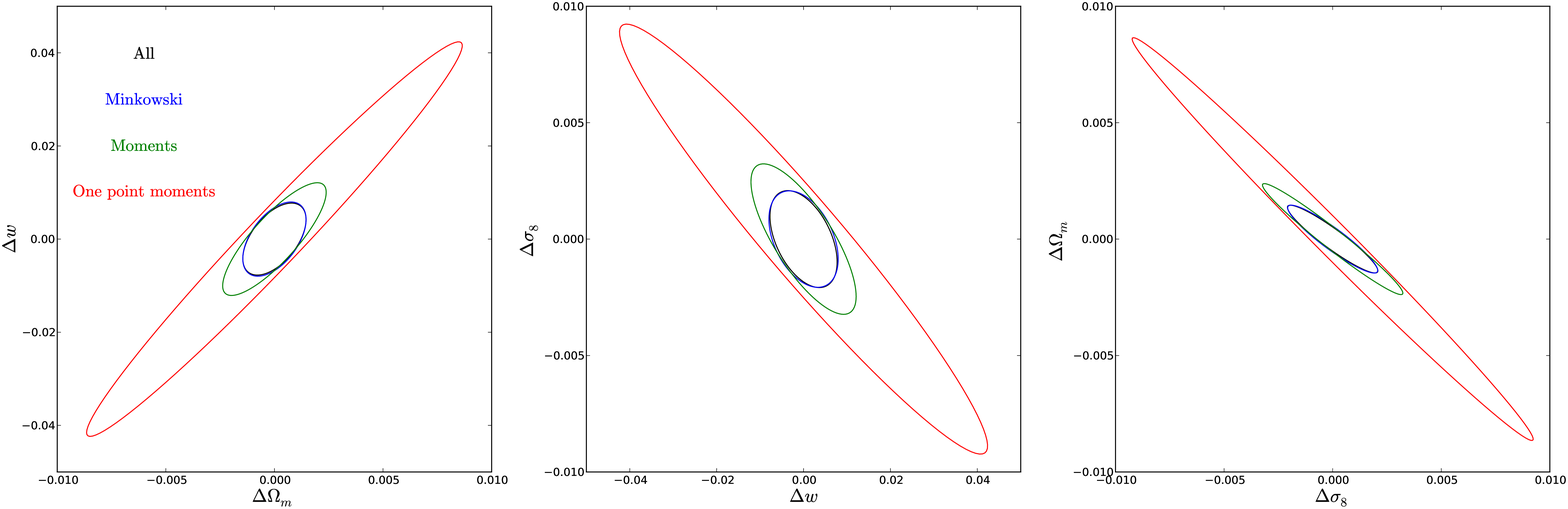}
\caption{This figure shows 68.4\% C.L. ellipses, drawn based on
  equation (\ref{chi2level}) with $n=2$. In the top panel, each point is the
  best-fit cosmology for one of the 1000 individual realizations, and
  the ellipse is plotted using the whole set of descriptors (including
  all three MFs and all moments up to kurtosis). 
  The bottom panel shows how these contours change with the type of
  descriptors used to fit the maps: MFs (blue), all
  moments up to kurtosis (green), one-point moments
  $(\sigma_0^2,S_0,K_0)$ only (red), and MFs together with moments
  (black).}
\label{ellipses}
\end{center}
\end{figure*}

We have investigated the errors in Table \ref{momcontributions} and checked if
these numbers do correspond to a 68.4\%, or one $\sigma$, confidence level.  Significant differences can
arise if the parameter fluctuations between realizations are not
Gaussian.  To quantify this better, and to obtain a more accurate
estimate of the $68.4\%$ confidence errors, we use the full
distribution of best-fit parameters.  If this distribution is Gaussian, then the quantity
\begin{equation}
\label{chi2withcov}
\chi^2_r=\delta p_\alpha^r(P^{AB})^{-1}_{\alpha\beta}\delta p_\beta^r
\end{equation}
should have a $\chi^2$ distribution with $n$ degrees of freedom, where $n$ is the number of dimensions, or number of parameters that we include in the statistics. If $n=1$ the points are distributed on a line, if $n=2$ on a two dimensional tilted ellipse, and if $n=3$ on a full three dimensional ellipsoid. For a Gaussian parameter distribution the constant $\chi^2$ surfaces are ellipses of equation
\begin{equation}
\label{chi2level}
\delta p_\alpha (P^{AB})^{-1}_{\alpha\beta}\delta p_\beta=k_n
\end{equation}
and the constant $k_n$, for a 68.4\% probability level, takes the values $k_1=1$, $k_2=2.3$ and $k_3=3.53$. In Table \ref{comptable} we compare the generalized Fisher marginalized errors computed from equation (\ref{covanalyticalmod}) with the 68.4\% confidence interval computed cutting the one dimensional parameter distribution at the edges. Since we have a good reason to believe that the errors are almost Gaussian, we use equation (\ref{chi2level}) with $n=2$ to plot the two dimensional confidence ellipses, that give us an idea about the error correlations between different parameters.These results are outlined in Figure \ref{ellipses}.
\begin{table*}
\begin{center}
\begin{tabular}{|c|c|c|c|} \hline
& $\Delta \Omega_m$& $\Delta w$& $\Delta \sigma_8$ \\ \hline
\multicolumn{4}{|c|}{\textbf{Minkowski} $V_0+V_1+V_2$} \\ \hline
Gaussian ($\sqrt{P_{\alpha\alpha}}$) & 0.00096 & 0.0052 & 0.0014\\
68.4\% interval & -0.0010$\div$0.00092&-0.0054$\div$0.0055 & -0.0013 $\div$ 0.0015 \\ \hline
\multicolumn{4}{|c|}{\textbf{One point moments} $(\langle\kappa^2\rangle,\langle\kappa^3\rangle,\langle\kappa^4\rangle_c$)} \\ \hline
Gaussian ($\sqrt{P_{\alpha\alpha}}$) & 0.0051 & 0.025 & 0.0054\\
68.4\% interval & -0.0028$\div$0.0037 & -0.016$\div$0.020&-0.0037$\div$0.0027 \\ \hline
\multicolumn{4}{|c|}{\textbf{All moments}} \\ \hline
Gaussian ($\sqrt{P_{\alpha\alpha}}$) & 0.0015 & 0.0081 & 0.0020\\
68.4\% interval & -0.0014$\div$0.0014&-0.0076$\div$0.0079 &-0.0020$\div$0.0019 \\ \hline
\multicolumn{4}{|c|}{\textbf{Minkowski + all moments}} \\ \hline
Gaussian ($\sqrt{P_{\alpha\alpha}}$) & 0.00096 &  0.0051 &  0.0014\\
68.4\% interval &-0.00099$\div$0.00091 &-0.0052$\div$0.0051 &-0.0013$\div$0.0014 \\ \hline
\end{tabular}
\caption{Comparison between marginalized errors calculated from the generalized Fisher matrix (\ref{covanalyticalmod}) and from the 68.4\% Monte Carlo confidence interval.}
\label{comptable}
\end{center}
\end{table*}
Table~\ref{comptable} shows that the analytical errors are
comparable with the ones calculated from the analytical
fitting. This feature, as stated above, is consistent with the
Gaussian nature of the errors.


\section{Discussion}
\label{discussion}
In this section we discuss our results and their implications.  
\subsection{Numerical accuracy of Minkowski functionals} 
From Figure \ref{gausscomp} we can see that, in the purely Gaussian
case, our code measures the MFs with an accuracy of
one part in $10^4$ for $V_0$ and one part in $10^3$ for the other
two. Even if this accuracy is not as good as that in \citep{MinkJan},
we can safely assert that we are able to establish whether the
perturbation series converges or not. As one can see in Figure
\ref{comparepert}, in the $\theta_G=1^\prime$ case, the perturbation
residuals are of relative order $10^{-2}$ for $V_0$ and $10^{-1}$ for
the other two. On the larger smoothing scales shown in
Figure~\ref{comparepertbetter}, the residuals are smaller, and become comparable with the code accuracy shown
in Figure~\ref{gausscomp}, suggesting that in fact the perturbation series is converging when the smoothing scale is $15^\prime$.
\subsection{Convergence of the perturbation series}  
Our main results, comparing the numerically measured MFs with the
approximations from the perturbative expansions, are shown in Figures
\ref{comparepert}, \ref{comparepertbetter} and \ref{chi2convergence}. While Figures \ref{comparepert} and \ref{comparepertbetter} show qualitatively when the perturbation series is a good approximation, Figure \ref{chi2convergence} suggests a quantitative criterion that helps deciding if the convergence is good or not: plotting the $\Delta \chi^2$ differences between measured and perturbative expanded MFs, we can see if the series converges or not looking at how this $\Delta \chi^2$ changes as we vary the perturbative order. We can see that on smoothing scales below $5^\prime$ the series diverges, it starts to converge on scales of $\sim 5^\prime$ and it shows a good degree of convergence on scales of $15^\prime$. We note that the galaxy shape noise in general favors the convergence of the series. 
This can be easily understood: since this noise is Gaussian in nature,
it helps reducing the non-Gaussianity in the map, and hence suppresses
the higher-order connected cumulants.  Nevertheless, these smoothing
scales are large, and much of the cosmological distinguishing power is
lost once the maps are smoothed on these scales. Looking at the
$\Delta \chi^2$ values in Table \ref{chi2}, we see that going from
$1^\prime$ to $5^\prime$, the distinguishing power measured by $\Delta
\chi^2$ drops by a factor of 3-10 in the noiseless case and 2-3 in the
noisy case.

\subsection{Cosmological Constraints}  
It is interesting to examine Table \ref{momcontributions}, where we
analyze the amount of information (or constraining power) carried by various subsets of
descriptors. The first important feature revealed in this table is
that, for the sake of constraining the triplet
$(\Omega_m,w,\sigma_8)$, the MFs lead to error-bars that are approximately 1.5$\div$2 times smaller than the ones obtained
with the multi-point moments that describe the
perturbation series at order $\sigma_0^2$.

We also investigated how the cosmological
information is distributed between these different moments.  One can
get interesting constraints just considering the three one point moments $(\sigma_0^2,S_0,K_0)$ -- i.e. just 3 numbers.  However, these
constraints are improved by a factor of 3 once we start considering
spatial information.  Most strikingly, most of the improvement can be
achieved by simply measuring $\sigma_1^2$ in addition to the one-point
moment. This is seen by going from the 2nd to the 3rd row of Table
\ref{momcontributions}; the addition of further multi-point moments
(4rd-8th rows) only tightens the errors further by a modest amount
($\sim10\%$).  From Table \ref{momcontributions}, it is also evident
that instead of measuring multi-point moments, spatial information can
be added by combining two or more smoothing scales (see \citep{MinkJan});
the similarly large benefits of combining smoothing scales have been
demonstrated for the lensing peak statistics \citep{Marian+2012}).
In fact, the table shows that combining just two smoothing scales
($1^\prime$ and $3^\prime$) has almost the same improvement, over the
one-point moments, as the addition of the multi-point moments, and the two agree even better if
we add a third ($1^\prime$,$3^\prime$ and $5^\prime$).   
The conclusion from the above is that the moments alone give constraints on the cosmological parameters that are 1.5$\div$2 worse than the MF ones,
even when spatial information is included, either in the form of
multi--point moments, or multiple smoothing scales. 
Another interesting possibility, which we didn't consider in this paper, is measuring cross correlators of the convergence field at different smoothing scales (i.e. quantities in the form $\langle\kappa_{\theta_1}\kappa_{\theta_2}...\kappa_{\theta_N}\rangle$). Even if we suspect that adding such correlators is equivalent to considering additional, intermediate smoothing scales in our analysis, the effect of these statistics on the parameter constraints is still not clear, and needs to be investigated in future work.
Finally, we see that all three MFs alone give comparable constraints, although as expected they carry some amount of complementary information as seen in the MFs section of Table \ref{momcontributions}. 
Since the moments do not seem to carry additional information once they are added to the MFs we investigated how actually these two sets of observables are correlated, measuring the cross correlation (averaging over realizations)
\begin{equation}
\gamma(p) = \frac{\langle\Delta p_{MF}\Delta p_{mom}\rangle}{\sqrt{\langle\Delta p_{MF}^2\rangle\langle\Delta p_{mom}^2\rangle}}
\end{equation}
which we measured to be $\gamma(\Omega_m)=0.34$, $\gamma(w)=0.35$ and $\gamma(\sigma_8)=0.45$; this shows that the moments do not add significant information to the one already contained in the MF even if these two descriptors are weakly correlated. We wish to make a final remark on the nature of the errors that we
forecast: if we look at Table \ref{comptable} we see that the
marginalized errors computed drawing the 68.4\% confidence ellipses
are comparable to the ones calculated analytically
with the formula (\ref{covanalyticalmod}). This means that our errors are almost Gaussian. 

\subsection{Robustness of the Results}
\label{robustness}
\begin{figure*}
\begin{center}
\includegraphics[scale=0.3]{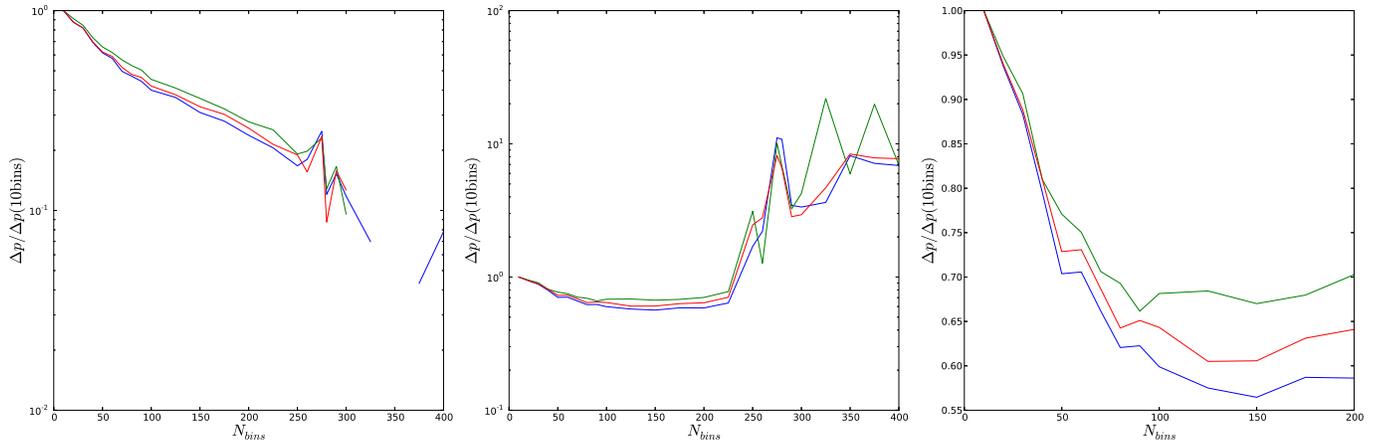}
\caption{Parameter constraints on $\Omega_m$(blue), $w$ (green) and $\sigma_8$ (red)  obtained varying the number of bins, using both equations (\ref{covanalytical}) in the left panel and (\ref{covanalyticalmod}) in the center panel. The right panel is obtained zooming the center panel on the plateau region $10<N_{bins}<200$; the breaks in the left panel correspond to cases in which the covariance matrix was close to singular and the constraint forecasts didn't converge. The errors are normalized to the ones that one obtains using $N_{bins}=10$.}
\label{numberbins}
\end{center}
\end{figure*}

\begin{figure*}
\begin{center}
\includegraphics[scale=0.3]{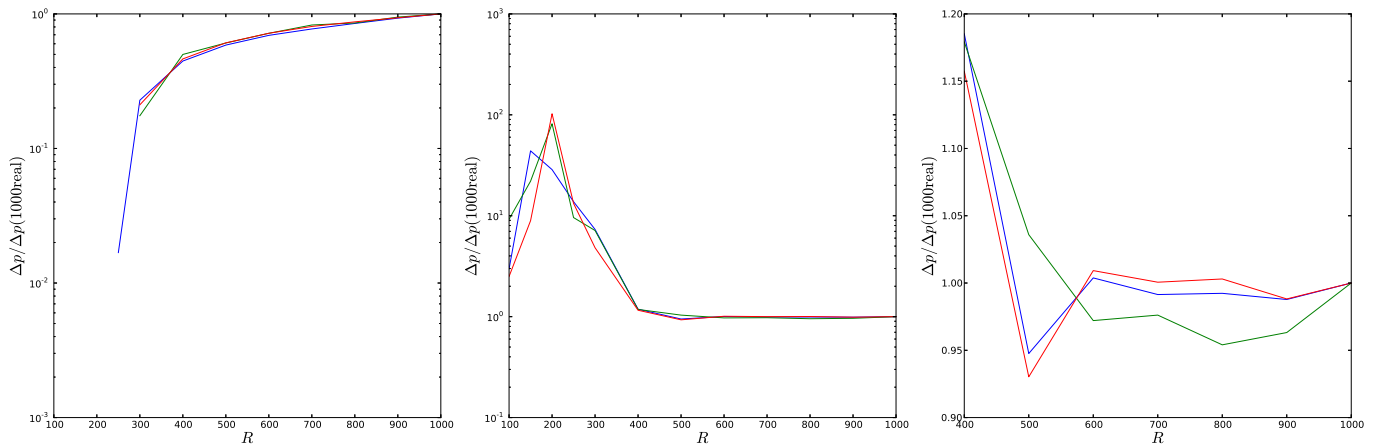}
\caption{Parameter constraints on $\Omega_m$(blue), $w$ (green) and $\sigma_8$ (red) obtained varying the number of realizations and keeping the number of bins fixed at $N_{bins}=100$. We use both equations (\ref{covanalytical}) in the left panel and (\ref{covanalyticalmod}) in the center panel. The right panel is obtained zooming the center panel on the plateau region $400<R<1000$; the breaks in the left panel correspond to cases in which the covariance matrix was close to singular and the constraint forecasts didn't converge. The errors are normalized to the ones that one obtains using $R=$1000 realizations.}
\label{nrealizations}
\end{center}
\end{figure*}

In this section we discuss some subtleties that can convince the reader of the robustness of our results.  First of all, we studied the effect of the choice of binning on our conclusions: in previous work \citep{MinkJan} have hypothesized that the errors calculated from equation (\ref{covanalytical}) could be underestimated if the covariance matrix $C_{ij}$ is too noisy, due to having too few realizations (see also \citep{Dodelson13}). We tried to avoid this problem generalizing equation (\ref{covanalytical}) with (\ref{covanalyticalmod}) taking advantage of the additional map set generated from 45 simulations. 
To check if this generalization actually helps in solving the problem, in Figures \ref{numberbins},\ref{nrealizations} we plotted the parameter errors $\Delta p_{\alpha}$ obtained from the MFs as a function of the number of bins $N_{bins}$ and number of realizations $R$ used. 

\begin{table}
\begin{center}
\begin{tabular}{|c|c|c|} \hline
$\Delta \Omega_m$ &$\Delta w$ & $\Delta \sigma_8$ \\ \hline
\multicolumn{3}{|c|}{\textbf{10 bins}} \\ \hline
\multicolumn{3}{|c|}{From $P_{\alpha\beta}$ in equation (\ref{covanalytical})} \\ \hline
\,\, 0.0016 \,\, &  \,\, 0.0077 \,\,  & \,\, 0.0022 \,\,  \\ \hline
\multicolumn{3}{|c|}{From $P_{\alpha\beta}^{AB}$ in equation (\ref{covanalyticalmod})} \\ \hline
0.0016  & 0.0076  &  0.0022 \\ \hline
\multicolumn{3}{|c|}{\textbf{100 bins}} \\ \hline
\multicolumn{3}{|c|}{From $P_{\alpha\beta}$ in equation (\ref{covanalytical})} \\ \hline
\,\, 0.00063 \,\, &  \,\, 0.0035 \,\,  & \,\, 0.00091 \,\,  \\ \hline
\multicolumn{3}{|c|}{From $P_{\alpha\beta}^{AB}$ in equation (\ref{covanalyticalmod})} \\ \hline
0.00096  & 0.0052 &  0.0014 \\ \hline
\end{tabular}
\end{center}
\caption{Comparison between the MF constraints on the cosmological parameters, calculated from equations (\ref{covanalytical}) and (\ref{covanalyticalmod}) for two different binning choices,$N_{bins}=10,100$.}
\label{10binstable}
\end{table}

\begin{figure*}
\begin{center}
\includegraphics[scale=0.3]{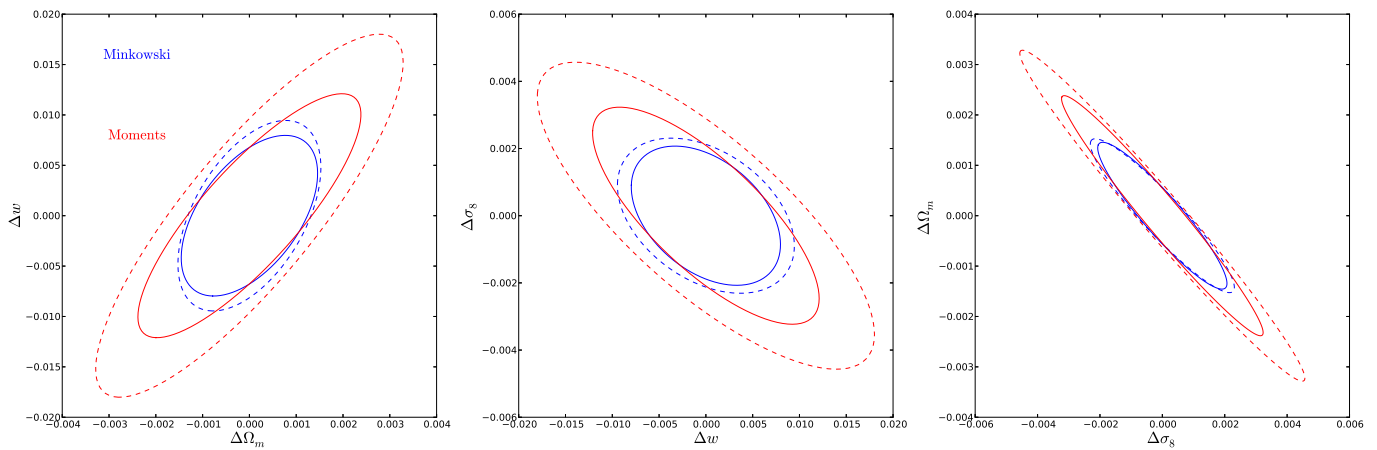}
\caption{Two dimensional ellipse contour plotted according to equation (\ref{chi2level}) using MFs (blue) and moments (red) as descriptors; in this figure we compare the effect of using forward (solid lines) and backward (dashed lines) derivatives in equation (\ref{interp}).}
\label{fwdvsbkw}
\end{center}
\end{figure*}

Even if Table \ref{10binstable} shows that using either equations (\ref{covanalytical}) or (\ref{covanalyticalmod}) does not make a big difference when the number of bins is small, it also shows that this is not the case when $N_{bins}\gtrsim 100$: in particular for $N_{bins}=100$ we see that the errors calculated following equation (\ref{covanalytical}) are about 1.5 smaller than the ones calculated according to equation (\ref{covanalyticalmod}). Figure \ref{numberbins} shows that when we use only one set of maps the constraints become artificially too small when we increase the number of bins from 10 to 400, mainly due to the fact that the covariance matrix becomes singular around $N_{bins}=300$ if we use 1000 realizations (see \citep{Dodelson13}). On the other hand, when we use different sets of maps, we see that the constraints reach a plateau in the region $100\lesssim N_{bins} \lesssim 200$ and start to blow up shortly after that due to numerical instabilities, as can be seen in the oscillating behaviour in the central panel of Figure \ref{numberbins}; this has already been observed in the study of peak statistics (\citep{PeaksJan}). We are confident that the constraints we obtain for $N_{bins}=100$ are realistic, because this value lies in the plateau region, where increasing the number of bins does not make a big difference. Moreover, we see in Figure \ref{nrealizations}, right panel that the constraints seem to converge to the actual values we find, once we increase the number of realizations, which makes us more confident about the constraints we obtain to be realistic; it is worth mentioning that in the right panel of Figure \ref{nrealizations} deviations from the true value due to too few realizations are towards larger, more conservative error bars, while in the left panel too few realizations cause error bars to be underestimated. 
Finally, we checked if using forward or backward derivatives in equation (\ref{interp}) affects our conclusions: to investigate this we plotted the two dimensional ellipse contours using equations (\ref{chi2level}) and (\ref{covanalyticalmod}) with $X$ calculated as backward derivatives, the results are displayed in Table \ref{fwdvsbkwtbl} and Figure \ref{fwdvsbkw}. 
\begin{table}
\begin{center}
\begin{tabular}{|c|c|c|} \hline
$\frac{\Delta \Omega_m\mathrm{(mom)}}{\Delta\Omega_m\mathrm{(MF)}}$ &$\frac{\Delta w\mathrm{(mom)}}{\Delta w\mathrm{(MF)}}$ & $\frac{\Delta \sigma_8\mathrm{(mom)}}{\Delta\sigma_8\mathrm{(MF)}}$ \\ \hline
\multicolumn{3}{|c|}{Forward derivatives} \\ \hline
1.6 & 1.5  & 1.6   \\ \hline
\multicolumn{3}{|c|}{Backward derivatives} \\ \hline
2.1 & 1.9  & 2.0 \\ \hline
\end{tabular}
\end{center}
\caption{Relative contributions of MFs and Moments to the $\Omega_m,w,\sigma_8$ error bars, computed with forward and backward derivatives, for $N_{bins}=100$}
\label{fwdvsbkwtbl}
\end{table}

We can see that, in the backward case, the parameter constraints due to MFs are a factor of 2 tighter than the ones due to moments; we compare this result with the one obtained using the forward derivatives, in which the comparison factor drops to 1.5. We suspect that the real case, where we go beyond the linear interpolation, will give a result somewhere in between, but the investigation of this requires a more advanced fitting procedure, because equation (\ref{tosolve}) becomes a non linear system. 
    
\section{Conclusions}
\label{conclusion}

In this work, we compared the amount of cosmological information in
MFs and multi-point moments, applied to weak lensing
convergence maps. We found that, even if a perturbative expansion of
the MFs in term of the moments is formally possible, this expansion
does not show a good degree of convergence until we smooth on scales
bigger than $\sim 5^\prime$. Unfortunately, these scales are so large that the majority
of the cosmological information in the convergence maps is lost, once
they are smoothed on these scales.

We have also shown that the moments of the convergence
fields, up to second order in the perturbation series, contain a
factor of $1.5-2$ less information than the full set of MFs. The three MFs alone carry comparable information, although one can improve the constraints combining the three MFs together.  
Regarding the information carried by moments, we have found that one can greatly improve the constraints on the
cosmological parameters once one adds spatial information - either by
considering low-order moments of spatial gradients of the field, or by
combining at least two different spatial smoothing scales.

In this work, we did not consider any systematic effects that have
instrumental (or atmospheric) origin. Although we added galaxy shape
noise to our maps, we did not consider the impact of imperfect
knowledge of this noise, or the impact of incomplete sky coverage,
atmospheric noise, PSF errors, etc. on our results.  It is also worth
noticing that the weak lensing convergence is not directly measurable
in actual experiments, and one needs to construct it from the $E$ mode
of the shear field. Since this construction is usually done in Fourier
space, we expect map masking to be important. To avoid Fourier space issues a possibility is using aperture mass filters, a method already widely used in the lensing community. 
Nevertheless, these issues will have to be addressed in future work.  Similarly, the impact of theoretical
systematic errors, such as uncertainties in baryon physics, or
intrinsic alignment of galaxies, will have to be studied in the
future.
We finally note then that the CFHTLenS survey \citep{CFHT13} has
recently measured the one point moments of the convergence field, up
to the fifth cumulant. Given the results we obtained in this work, we
suggest that measuring even one of the moments that carries spatial
information (i.e. one of the moments with derivatives), in addition to
the ones already measured, could improve the constraints on the
cosmological parameters significantly.

\newpage
\section*{Acknowledgements}

We thank Deepak Munshi and Kevin Huffenberger for useful discussions. 
We thank the referee for the insightful comments.  
This research utilized resources at the New York Center for Computational Sciences, a
cooperative effort between Brookhaven National Laboratory and Stony
Brook University, supported in part by the State of New York. This
work is supported in part by the U.S. Department of Energy under
contracts DE-AC02-98CH10886 and 
DE-FG02-92-ER40699, 
by the NSF under grant AST-1210877, and
by the NASA under grant NNX10AN14G.

\newpage

\bibliography{ref}
\label{lastpage}
\end{document}